\documentclass[12pt, prd, showpacs]{revtex4}
\usepackage{amssymb}
\usepackage{amsmath}

\input{tcilatex}

\begin{document}

\title{Ultra-high energy collisions of nonequatorial geodesic particles near
dirty black \ holes}
\author{Oleg B. Zaslavskii}
\affiliation{Kharkov V.N. Karazin National University, 4 Svoboda Square, Kharkov, 61077,
Ukraine}
\email{zaslav@ukr.net}

\begin{abstract}
We consider collision of two geodesic particles moving around rotating
stationary axially symmetric black holes. It is shown for arbitrary
nonequatorial motion that under certain conditions the energy in their
centre of mass frame can grow unbound (the so-called BSW effect). This
generalizes the previous results for equatorial motion around dirty
(surrounded by matter) black holes and nonequatorial motion around the Kerr
metric. It turns out that the BSW effect occurs near any point of the
horizon surface. We do not use special symmetries of space-time typical of
the Kerr metric, so the results are quite generic. The general scheme
classifying all possible scenarios is discussed.
\end{abstract}

\maketitle




\section{Introduction}

If two particles moving towards the horizon of a black hole collide, under
certain conditions their energy in the centre of mass (CM) frame can become
\ infinitely large. This interesting effect was discovered by Ba\~{n}ados,
Silk and West \cite{ban} (called the BSW effect after the names of its
authors) and is under active study now. It is of interest from the
theoretical viewpoint as nontrivial phenomenon in gravity and can have
potential astrophysical consequences. Although for the Kerr black hole \cite%
{p}, \cite{j} the products of such collisions have a quite modest energy in
the frame of a distant observer due to strong red shift, the observational
outcome can become more significant for dirty black holes \cite{z1}. In
addition, there are hopes on some indirect manifestations of this effect due
to new channels of reactions with transmutation of particles (in particular,
of dark matter) near the black hole horizon \cite{gp09}, \cite{n}.

As far as the properties of the BSW effect are concerned, one of the main
questions here is to what extent it is universal. In the original paper \cite%
{ban} the extremal Kerr metric was considered and it was assumed that
particles move in the equatorial plane. Later on, it was understood that the
similar effect reveals itself also for nonextremal black holes \cite{gp}.
The general picture was described and it was shown that the BSW effect
arises for generic dirty (surrounded by matter) rotating black holes due to
the properties of the horizon, so it can be viewed as a manifestation of
universality of black hole physics \cite{prd}. However, this feature was
traced for motion in the equatorial plane only thus somewhat restricting the
statement about the universality of the BSW effect. Meanwhile, for
nonequatorial motion direct collisions between particles near extremal Kerr
black hole also lead to the BSW effect but with another kind of restriction:
it was found to occur in the bounded belt centered near the equator \cite%
{neq}.

The aim of the present paper is to combine both factor and consider the BSW
effect for dirty black holes for nonequatorial motion of colliding geodesic
particles to generalize previous results \cite{prd} and \cite{neq}. The new
qualitative results consists in that we show that the BSW effect may occur
in the vicinity of any point of the horizon and in this sense it retains its
universality. There is no contradiction here with the results of \cite{neq}
since there are different types of the BSW effect. Their classification was
discussed in \cite{neq} for the Kerr metric and is now extended to dirty
black holes in the present work. In particular, the restriction to the belt
found in \cite{neq} concerns the BSW effect due to direct collisions whereas
the BSW effect in polar region requires multiple scattering similar to the
BSW effect for nonextremal black holes \cite{gp}, \cite{prd}.

The essential point of the analysis in \cite{neq} consisted in the fact that
the Kerr metric possesses a remarkable property - separation of variables in
the Hamilton - Jacobi equation \cite{car}. It does not hold for a generic
black hole space-time that does not allow the extend the approach of \cite%
{neq}.\ Meanwhile, we suggest a more simple general approach that does not
demand separability of variables and applies to an arbitrary axially
symmetric dirty black hole.

One reservation is order. We do not address in this paper an important issue
about the possibility to evade limitations on the energy detected at
infinity found earlier for equatorial motion \cite{p} - \cite{z1}. However,
the present work can serve as a basis for further investigation of this
issue.

\section{Basic formulas and limiting transitions}

Consider the generic axially symmetric metric.\ It can be written as%
\begin{equation}
ds^{2}=-N^{2}dt^{2}+g_{\phi \phi }(d\phi -\omega dt)^{2}+\frac{\rho ^{2}}{%
\Delta }dr^{2}+g_{\theta \theta }d\theta ^{2}.  \label{teta}
\end{equation}%
Here, the metric coefficients do not depend on $t$ and $\phi $. On the
horizon $N=0$. \ In (\ref{teta}), the factor $\Delta (r)\sim N^{2}$ is
singled out for convenience. The coefficient $\rho $ can depend on $\theta $%
. Near the horizon, $\Delta \sim N^{2}$, so $r$ is the analog of the
quasiglobal coordinate used in the spherically symmetric case \cite{br}. It
is worth noting that the form of the metric somewhat differs from that in
the Gauss normal coordinates used in \cite{v1}, \cite{v2}, \cite{tan}. It is
more convenient for our purposes and, in particular, facilitates the
comparison to the Kerr metric.

In the space-time under discussion there are two conserved quantities $%
E\equiv -mu_{0}$ and $L\equiv mu_{\phi }$ where $u^{\mu }=\frac{dx^{\mu }}{%
d\tau }$ is the four-velocity of a test particle having the mass $m$, $\tau $
is the proper time and $x^{\mu }=(t,\phi ,r,\theta )$ are coordinates..The
aforementioned conserved quantities have the physical meaning of the energy
(or frequency for a light-like particle) and azimuthal component of the
angular momentum, respectively. Then, using these first integrals for such
geodesics one can write down equation of motion (dot denotes the derivative
with respect to the proper time $\tau $):%
\begin{equation}
m\dot{t}=mu^{0}=\frac{X}{N^{2}}\text{, }X=E-\omega L\text{,.}  \label{t}
\end{equation}%
We assume that $\dot{t}>0$, so that $E-\omega L\geq 0$.%
\begin{equation}
m\dot{\phi}=\frac{L}{g}+\frac{\omega X}{N^{2}}\text{, }g=g_{\phi \phi }.
\label{phi}
\end{equation}%
\begin{equation}
\frac{\rho ^{2}}{\Delta }m^{2}\dot{r}^{2}=V_{eff}\equiv \frac{X^{2}}{N^{2}}-%
\frac{L^{2}}{g}-m^{2}-m^{2}g_{\theta \theta }\dot{\theta}^{2}\equiv \frac{%
Z^{2}}{N^{2}}.  \label{r}
\end{equation}%
Here, $V_{eff}=\frac{Z^{2}}{N^{2}}$ has the meaning of the effective
potential.

The quantity which is relevant for us is the energy in the centre of mass
frame $E_{c.m.}$ \cite{ban} where%
\begin{equation}
E_{c.m.}^{2}=-\left( m_{1}u_{1}^{\mu }+m_{2}u_{2}^{\mu }\right) \left(
m_{1}u_{1\mu }+m_{2}u_{2\mu }\right) \text{, }
\end{equation}%
subscript i=1,2 enumerates particles. After simple manipulations, one
obtains from (\ref{t}) - (\ref{r}) that%
\begin{equation}
E_{c.m.}^{2}=m_{1}^{2}+m_{2}^{2}+2m_{1}m_{2}\gamma \text{, }\gamma
=-u_{1}^{\mu }u_{2\mu }\text{,}  \label{ecm}
\end{equation}%
\begin{equation}
\gamma =c-d-g_{\theta \theta }\dot{\theta}_{1}\dot{\theta}_{2}\text{, }c=%
\frac{X_{1}X_{2}-Z_{1}Z_{2}\text{, }}{m_{1}m_{2}N^{2}}\text{, }d=\frac{%
L_{1}L_{2}}{m_{1}m_{2}g_{\phi \phi }}\text{.}  \label{ga}
\end{equation}%
As is known, the BSW effect arises when one of two colliding particle is
critical (near-critical) and the other one is usual. By definition, a
particle is usual if $X_{H}\neq 0$ and is critical if $X_{H}=0$, $E=\omega
_{H}L$ or near-critical if $X_{H}$ is small. Here, subscript "H" means that
a corresponding quantity is calculated on the horizon. Let particle 1 be
critical or near-critical and particle 2 be usual. It follows from (\ref{ecm}%
) that in the near-horizon region where $N\rightarrow 0$, $\left(
Z_{2}\right) _{H}\approx \left( X_{2}\right) _{H}$, 
\begin{equation}
E_{c.m.}^{2}\approx 2\frac{\left( X_{2}\right) _{H}}{N^{2}}(X_{1}-Z_{1})%
\text{.}  \label{e}
\end{equation}%
From now on, we consider two cases separately.

\section{Extremal black holes}

\subsection{Infinite growth of energy in the CM frame}

Let a particle be not exactly critical but near critical, so%
\begin{equation}
L=\frac{E}{\omega _{H}}(1-\delta )\text{, }\delta \ll 1\text{.}  \label{l}
\end{equation}

Then, 
\begin{equation}
X\approx E(1-\frac{\omega }{\omega _{H}}+\delta )\text{.}  \label{xd}
\end{equation}

Near the horizon of the extremal black hole, $\omega -\omega _{H}$ has the
order $N$ \cite{tan}, so%
\begin{equation}
\omega =\omega _{H}-B(\theta )N+O(N^{2})\text{.}  \label{w}
\end{equation}%
We also adjust $\delta $ to have the same order $N$, so%
\begin{equation}
\delta =CN+O(N^{2})\text{.}  \label{d}
\end{equation}%
In particular, one can choose $C=0$ as it was actually done in \cite{neq}.

Then, for the near-critical particle we have near the horizon%
\begin{equation}
X\approx \frac{E}{\omega _{H}}\tilde{B}N\text{, }\tilde{B}=B+C\omega _{H}.
\label{x}
\end{equation}%
It is seen that the right hand side of (\ref{r}) is finite for such a
particle,%
\begin{equation}
\left( V_{eff}\right) _{H}\approx \alpha -m^{2}g_{\theta \theta }\dot{\theta}%
^{2}\text{, }\alpha \equiv \frac{E^{2}}{\omega _{H}^{2}}(\tilde{B}_{H}^{2}-%
\frac{1}{g_{H}})-m^{2}\text{.}  \label{a}
\end{equation}%
Then, it follows from (\ref{e}) that%
\begin{equation}
E_{c.m.}^{2}\approx 2\frac{\left( X_{2}\right) _{H}}{N}\beta .  \label{be}
\end{equation}%
\begin{equation}
\beta =\frac{E_{1}\tilde{B}}{\omega _{H}}-\sqrt{\alpha
_{1}-m_{1}^{2}g_{\theta \theta }\dot{\theta}_{1}^{2}}\text{,}  \label{b}
\end{equation}%
it is assumed that particle 1 is near-critical, particle 2 is usual.

In the limit $N\rightarrow 0$ the CM energy grows unbound: $E_{c.m.}\sim
N^{-1/2}$ similar to the equatorial motion case \cite{prd}, so the BSW
effect takes place.

\subsection{Kinematic conditions}

This is not the end of story since for the realization of the BSW it is
necessary that a critical particle approach the horizon. This gives rise to
the condition $Z_{1}^{2}\geq 0$ that entails 
\begin{equation}
\alpha _{1}\geq 0,  \label{al}
\end{equation}%
whence%
\begin{equation}
C\geq \eta (\theta )\equiv \sqrt{\frac{1}{\omega _{H}^{2}g_{H}}+\frac{%
m_{1}^{2}}{E_{1}^{2}}}-\frac{B}{\omega _{H}}\text{.}  \label{c}
\end{equation}

The value of $\theta $ where $\alpha _{1}=0$ (which is equivalent to $C=\eta
(\theta )$) is just the turning point for a variable $\theta $. In the case
of equatorial motion $\theta =\frac{\pi }{2}=const$ the term $%
m_{1}^{2}g_{\theta \theta }\dot{\theta}_{1}^{2}$ in (\ref{a}) is identically
zero. If $\eta \leq 0$ in some interval of $\theta $, one can simply put
there $C=0.$ The corresponding region generalizes the belt obtained in \cite%
{neq} for critical particles. Inside this region, direct collisions lead to
the BSW effect since particle 1 is exactly critical. Outside this belt, we
have $\eta >0$ and the critical particle ($\delta =C=0$) cannot reach the
horizon since condition (\ref{al}) is violated in its vicinity. In the
region $\eta >0$, a particle should be near-critical (not exactly critical)
with $C$ satisfying eq. (\ref{c}). Here, the quantities $g_{H}$ and $B$
depend, in general, on $\theta $. Therefore, if one wants to arrange
collision between particles near the point on the horizon with some value $%
\theta ^{\ast }$ of the polar angle, the corresponding minimum value of $C$
also depends on $\theta ^{\ast }$ according to (\ref{c})$.$ If $m\ll E$,
condition (\ref{c}) simplifies to $C\omega _{H}\geq \frac{1}{\sqrt{g_{H}}}-B$%
.

Near the polar axis, in the absence of the conical defect $g\sim \theta ^{2},
$ so in the limit $\theta \rightarrow 0$, we obtain that the admissible $C$
grows like $\frac{1}{\theta }$. To be consistent with the condition (\ref{l}%
), collision should occur so closely to the horizon that $N\ll A\theta $
where $A~$is a constant. Then, it follows from (\ref{a}) - (\ref{b}) that $%
E_{c.m.}^{2}$ is proportional to $\delta ^{-1}$.

To realize the BSW effect in the region forbidden for the critical particle
near the horizon, the scenario of multiple scattering \cite{gp} can be used.
A\ usual particle can approach the horizon and, near it, get a near-critical
value of the angular momentum in some collision and only after that collide
one more time with a usual particle thus producing the BSW effect.

\section{Extremal Kerr metric}

Let us consider the Kerr metric. In the Boyer-Lindquist coordinates \cite{bl}%
, it can be written as%
\begin{equation}
ds^{2}=-dt^{2}(1-\frac{2Mr}{\rho ^{2}})-\frac{4Mar\sin ^{2}\theta }{\rho ^{2}%
}d\phi dt+\frac{\rho ^{2}}{\Delta }dr^{2}+\rho ^{2}d\theta ^{2}+(r^{2}+a^{2}+%
\frac{2Mra^{2}\sin ^{2}\theta }{\rho ^{2}})\sin ^{2}\theta d\phi ^{2}\text{,}
\label{k}
\end{equation}%
where $\rho ^{2}=r^{2}+a^{2}\cos ^{2}\theta $, $\Delta =r^{2}-2Mr+a^{2}$, $M$
is the black hole mass, $a$ characterizes its angular momentum. It follows
from (\ref{k}) that

\begin{equation}
\omega =\frac{2aMr}{(r^{2}+a^{2})^{2}-a^{2}\Delta \sin ^{2}\theta }\text{,}
\end{equation}%
\begin{equation}
N^{2}=\frac{\Delta \rho ^{2}}{(r^{2}+a^{2})^{2}-a^{2}\Delta \sin ^{2}\theta }%
\text{.}
\end{equation}%
For the extremal horizon, $M=a$, and comparing the exact expressions with
the near-horizon expansion (\ref{w}) one can find easily that 
\begin{equation}
\omega _{H}=\frac{1}{2M}\text{, }B=\frac{1}{M\sqrt{1+\cos ^{2}\theta }}\text{%
, }g_{H}=4\frac{M^{2}\sin ^{2}\theta }{1+\cos ^{2}\theta }\text{.}
\end{equation}%
Then, (\ref{c}) gives us%
\begin{equation}
C\sqrt{1+\cos ^{2}\theta }\geq \sqrt{\frac{(1+\cos ^{2}\theta )^{2}}{\sin
^{2}\theta }+\frac{m_{1}^{2}(1+\cos ^{2}\theta )}{E_{1}^{2}}}-2\text{.}
\label{ck}
\end{equation}

If $C=0$, (\ref{ck}) takes the form%
\begin{equation}
\left( m_{1}^{2}-E_{1}^{2}\right) \sin ^{4}\theta
+2(4E_{1}^{2}-m_{1}^{2})\sin ^{2}\theta -4E_{1}^{2}\geq 0  \label{0k}
\end{equation}%
that coincides exactly with eq. (4.5) of Ref. \cite{neq}, so further
analysis developed in \cite{neq} applies. The appearance of the belt
restricting the region of the BSW effect follows just from (\ref{0k}).
However, if one adjusts $C\neq 0$ that satisfies (\ref{ck}), this effect can
occur near any polar angle including the region forbidden for pure critical
particles.

\section{Nonextremal black holes}

For the nonextremal horizon, in the metric coefficient $\omega $ the
correction to its horizon value $\omega _{H}$ has the order $N^{2}$ \cite{v2}%
, \cite{tan}. Therefore, for the critical particle $X^{2}\sim N^{4}\ll N^{2}$%
, so in (\ref{r}) $Z_{1}^{2}<0$ which means that such a particle cannot
reach the horizon in agreement with previous observations \cite{prd}, \cite%
{ne}. Let a particle be near-critical with $\delta $ having the form (\ref{d}%
). Then, neglecting in (\ref{xd}) the term of the order $N^{2}$ that comes
from $\omega -\omega _{H}$, we obtain 
\begin{equation}
X\approx ECN\text{.}
\end{equation}%
instead of (\ref{x}).

The general expression (\ref{be}) holds with%
\begin{equation}
\beta =E_{1}C-\sqrt{\alpha -m^{2}g_{\theta \theta }\dot{\theta}_{i}^{2}}%
\text{,}
\end{equation}%
\begin{equation}
\alpha =E^{2}(C^{2}-\frac{1}{\omega _{H}^{2}g_{H}})-m^{2}.
\end{equation}

The restriction on $C$ takes the form%
\begin{equation}
C\geq \sqrt{\frac{m^{2}}{E^{2}}+\frac{1}{\omega _{H}^{2}g_{H}(\theta )}}%
\text{.}  \label{c2}
\end{equation}

The inequality (\ref{c2}) generalizes the restriction derived in eq. (18) of
Ref. \cite{prd} where it was assumed that $\theta =\frac{\pi }{2}$ (which,
in turn, generalized eq. (18) of \cite{gp}).

\section{Different scenarios of the BSW effect}

In \cite{neq} classification of different scenarios of collisions leading to
the BSW effect was suggested.\ It is based on behavior of the effective
radial potential for motion of critical particles in the vicinity of the
horizon and takes into account the type of the horizon (extremal or
nonextremal). The possibility to single out pure radial motion is based on
the fact that variables in the Hamilton - Jacobi equation are separated for
the Kerr metric. Now, for generic dirty black holes this is, generally
speaking, not so. Nonetheless, it follows from the above consideration that
this scheme is extendable to the general case of dirty black holes. In doing
so, the relevant quantity that replaces the Boyer-Lindquist coordinate $r$
used in \cite{neq} for the Kerr metric is the lapse function $N$. For
nonextremal black holes $N^{2}\sim r-r_{H}$, for extremal ones $N\sim
r-r_{H} $. Then, one obtains the scenario of type I if near the horizon $%
Z^{2}\approx AN^{2}$, $A>0$ (with the extremal horizon), type II if $A=0$,
type III if $A<0$ (with the extremal horizon), type IV if $A<0$ with a
nonextremal horizon.

Type I corresponds to direct collisions since the potential has the correct
sign near the horizon, so a critical particle can safely reach it. The BSW
process considered in the pioneering work \cite{ban} belongs just to this
type. Type II corresponds to circular orbits. The BSW effect due to
collisions of particles on such orbits was considered in \cite{kerr} for the
Kerr metric and in \cite{circ} for dirty black holes. Type III was mentioned
in \cite{neq} as the case not discussed in literature before. Meanwhile, in
our context, type III is especially interesting since one can recognize here
just the BSW effect in polar regions for extremal horizon that is the main
subject of the present paper! Type IV is generalization of scenario
considered in \cite{gp} for the equatorial motion in the Kerr space-time.
Formally, one can also consider type V: nonextremal horizons with if $A>0.$
It was mentioned in \cite{neq} but not included in the table since it cannot
be realized for the Kerr metric. Meanwhile, it is clear from the above
consideration that it cannot be realized in general as well. Indeed, the
correct sign of $Z^{2}\,\ $would mean that the critical particle can reach a
nonextremal horizon. This is impossible, as is explained in the previous
section.

\section{Summary and conclusions}

Thus we considered collision of geodesic particles for arbitrary
nonequatorial motion and showed that the BSW effect occurs in the vicinity
of generic dirty rotating axially symmetric stationary black holes. This
happens not only inside some parts of the horizon surface but near an
arbitrary point on it. These results fill some gaps in earlier results and
makes the picture complete. All possible scenarios are united in a scheme
that generalizes the previous results for the Kerr metric. It turns out that
the BSW effect reveals itself irrespective of the presence or absence of
special properties of a space-time like separability variables of variables
in the Hamilton - Jacobi equation.

Meanwhile, it should be mentioned that our consideration is based on a
simplified picture of motion along geodesics. In more realistic
circumstances, one should also take into account additional effects like
gravitational radiation \cite{ted}, \cite{berti}, synchrotron radiation of
charged particles in the magnetic field \cite{fr}, etc. Then, the whole
picture can drastically change and the key question is whether these effects
can set a limit to the energy that can be reached. At present, the answer is
not obvious since there are indications that the BSW effect retains its
validity (at least, for neutral particles) provided there are critical
trajectories of general character, even if they are not geodesics \cite{gc}.
Further careful investigation is needed here.

\begin{acknowledgments}
This work was supported in part by the Cosmomicrophysics section of the
Programme of the Space Research of the National Academy of Sciences of
Ukraine.
\end{acknowledgments}

\end{document}